\documentstyle[aps,12pt,manuscript]{revtex}
\begin{document}
\preprint{INJE--TP--95--5 }
\def\overlay#1#2{\setbox0=\hbox{#1}\setbox1=\hbox to \wd0{\hss #2\hss}#1%
\hskip -2\wd0\copy1}

\title{ Instability of  a two-dimensional extremal black hole}

\author{  Y. S. Myung }
\address{Department of Physics, Inje University, Kimhae 621-749, Korea}

\author{ Jin Young Kim}
\address{Division of Basic Science, Dongseo University, Pusan 616-010, Korea}

\maketitle
\vskip 1.5in

\begin{abstract}
We consider the perturbation of tachyon about the extremal ground state of
a two-dimensional (2D)
electrically charged black hole.
It is found that the presenting potential to on-coming tachyonic wave takes a
double-humped barrier well. This allows an exponentially growing mode with
respect to time.
This extremal ground state  is
classically unstable. We conclude that the 2D extremal electrically charged
black hole cannot be
 a candidate for the stable endpoint of the Hawking evaporation.

\end{abstract}

\newpage
Recently the extremal black holes have received much attention.  Extremal
black hole provide
a simple laboratory in which to investigate the quantum aspects of black
hole [1]. One of the crucial
features is that the Hawking temperature vanishes.  The extremal
magnetically charged black holes are
shown to be classically as well as quantum mechanically stable [2]. The
black hole with $M>Q$
  will tend to Hawking
radiate down to its extremal $M=Q$ state. Thus the extremal black hole may
play  a role of
the stable endpoint for the Hawking evaporation. It has been more recently
proposed that although
the extremal black hole has nonzero area, it has zero entropy [3]. This is
because the extremal case is
distinct topologically  from the nonextremal one. For example, the extremal
black hole has
the infinite throat. The origin is effectively removed from the manifold.
The topology is no longer a disk  but
rather, an annulus whose inner boundary is at infinite distance.

In this letter, we will address the  stability aspects of the 2D extremal
electrically charged
 black hole.
It  is the fundamental test for the 2D extremal black hole to exist.
We always  visualize the
black hole as presenting an effective potential barrier (or well) to the
on-coming waves [4].
One easy way of understanding the attributes of an extremal black hole is
to find
out how it reacts to external perturbations.
In deciding whether or not the the extremal black hole is stable, one
starts with a physical
perturbation which is regular everywhere in space at the initial time $t=0$
[5]. And then see whether such a
perturbation will grow with time. If there exists an exponentially growing
mode, the extremal
black hole is unstable.

We will work with part of the low energy action to heterotic string theory [6]
\begin{equation}
S_{l-e} = \int d^2 x \sqrt{-G} e^{-2\Phi}
   \big \{ R + 4 (\nabla \Phi)^2 + \alpha^2 - {1 \over 2} F^2 - {1 \over 2}
(\nabla T)^2 +
 T^2 \big \}.
\end{equation}
Here are all string fields ( metric $G_{\mu\nu}$,  dilaton $\Phi$,  Maxwell
field $F_{\mu\nu}$,
 and tachyon $T$).
Setting $\alpha^2 = 8$ and  after deriving equations, we take the
transformation
\begin{equation}
-2\Phi \rightarrow \Phi,~~~ T \rightarrow \sqrt 2 T, ~~~-R  \rightarrow R.
\end{equation}
Then the equations of motion become
\begin{eqnarray}
&&R_{\mu\nu} + \nabla_\mu \nabla_\nu \Phi + F_{\mu\rho}F_{\nu}^{~\rho} +
\nabla_\mu T \nabla_\nu T = 0,  \\
&& (\nabla \Phi)^2 +  \nabla^2 \Phi - {1 \over 2} F^2  - 2 T^2 - 8 = 0,  \\
&&\nabla_\mu F^{\mu \nu} + (\nabla_\mu \Phi) F^{\mu \nu} = 0,   \\
&&\nabla^2 T + \nabla \Phi \nabla T + 2 T = 0.
\end{eqnarray}

An electrically charged black hole solution to the above equations is given by

\begin{equation}
\bar \Phi = 2 \sqrt 2 r,~~~ \bar F_{tr} = Q e^{-2 \sqrt 2 r},~~~ \bar T = 0,
{}~~~ \bar G_{\mu\nu} =
 \left(  \begin{array}{cc} - f & 0  \\
                             0 & f^{-1}   \end{array}   \right),
\end{equation}
with
\begin{equation}
f = 1 -  {2M \over 2\sqrt 2}e^{- 2 \sqrt 2 r} + {Q^2 \over 8}e^{- 4 \sqrt 2 r},
\end{equation}
where $M$ and $Q$ are the mass and charge of the black hole, respectively.
For convenience, we take $ M=\sqrt2$.    For $0<|Q|<M$, the double horizons
($r_{\pm}$) are given by
\begin{equation}
r_{\pm} = { 1 \over 2 \sqrt 2 }\log \left[ {1 \pm  \sqrt { 1 - {Q^2 \over
2}} \over 2}\right],
\end{equation}
where $r_{+}(r_{-})$ correspond to the event (Cauchy) horizons.
This charged black hole may provide an ideal setting for studying the late
stages
of Hawking evaporation. For $Q=M$, two horizons coincide :$
r_{+}=r_{-}\equiv r_o$.
We are here interested  in this extremal limit.

To study the propagation of string fields, we introduce small perturbation
fields  around
the background solution as [7]
\begin{eqnarray}
&&F_{tr} = \bar F_{tr} + {\cal F}_{tr} = \bar F_{tr} [1 - {{\cal F}(r,t)
\over Q}],        \\
&&\Phi = \bar \Phi + \phi(r,t),                       \\
&&G_{\mu\nu} = \bar G_{\mu\nu} + h_{\mu\nu}  = \bar G_{\mu\nu} [1 - h
(r,t)],     \\
&&T = \bar T + \tilde t \equiv \exp (-{\bar \Phi \over 2}) [ 0 + t (r,t) ].
\end{eqnarray}
 One has to linearize (3)-(6) in order to obtain the equations governing
the perturbations.
It is important to check whether the graviton ($h$),  dilaton ($\phi$), Maxwell
mode (${\cal F}$)  and tachyon ($t$) are  physically propagating modes
in the 2D charged black hole background.
We review the conventional counting of degrees of freedom.
The number of degrees of freedom for the gravitational field ($h_{\mu\nu}$) in
$D$-dimensions is $(1/2) D (D -3)$.  For a Schwarzschild black hole,
we obtain two degrees of freedom. These correspond to the Regge-Wheeler
mode for odd-parity perturbation
and Zerilli mode for even-parity perturbation [4].  We have $-1$ for $D=2$.
This means that in
two dimensions
the contribution of the graviton is equal and opposite to that of a
spinless particle (dilaton).
The graviton-dilaton modes ($h+\phi, h-\phi$) are gauge degrees of freedom
and thus turn out to be
nonpropagating modes[6].
In addition, the Maxwell field has $D-2$ physical degrees of freedom.
The Maxwell field has no physical degrees of freedom for $D=2$. Actually
from (5) it turns out to be
a redundant one $( {\cal F}= -Q(h+\phi))$.
Since these all are  nonpropagating modes, it is not necessary to linearize
(3)-(4).
The stability should be based on the physical degrees of freedom.
The tachyon is a physically propagating mode. Its linearized equation is
\begin{equation}
f^2 t''  +  ff't'- [\sqrt 2 f f' - 2 f (1-f)]t  - \partial_t^2 t = 0,
\end{equation}
where the prime ($\prime$) denotes the derivative with respect to $r$.
To study the stability,  the above equation
should be transformed into the  one-dimensional Schr\"odinger equation.
Introducing the  coordinate transformation
$$r\to r^* \equiv g(r),$$
(14) can be rewritten as
\begin{equation}
f^2 g'^2 {\partial^2 \over \partial r^{*2}} t  + f \{ f g'' +  f' g'\}
{\partial \over \partial r^* }t - [\sqrt 2 ff' - 2 f (1 - f)]t
 - {\partial^2 \over \partial t^2} t = 0.
\end{equation}
Requiring that the coefficient of the linear derivative vanish, one finds
the relation
\begin{equation}
g' =  {1 \over f}.
\end{equation}
Assuming $t( r^*,t ) \sim \tilde t ( r^* ) e^{i\omega t}$,
one can cast (15) into  one-dimensional Schr\"odinger equation

\begin{equation}
\{ {d^2 \over dr^{*2}} + \omega^2 - V(r)\} \tilde t = 0,
\end{equation}
where the effective potential $V(r)$  for the extremal case ($Q=M=\sqrt 2$)
is given by

\begin{equation}
V(r) = f(\sqrt 2 f' - 2  (1 - f))
     = 2 e^{- 2 \sqrt 2 r} ( 1- {1 \over 2}e^{- 2 \sqrt 2 r})^2 ( 1- {3
\over 4}e^{- 2 \sqrt 2 r}) .
\end{equation}
The event horizon is located at $ r_o= -0.245$.
As is shown in Fig.1, we have a double-humped barrier well outside the
black hole.
According to the analysis of potentials, a double-humped potential appears
when the nonextremal
black hole (a simple barrier) approaches the extremal one.
Now let us translate the potential $V(r)$ into $V(r^*)$.
With $f= (1- {1 \over 2} \exp(-2\sqrt 2 r))^2$, one obtain the explicit
form of $r^*$
\begin{equation}
r^*= r - {1 \over 2\sqrt 2 (1- {1 \over 2} e^{-2\sqrt 2 r})} +
    {1 \over 2 \sqrt 2} \log |1- {1 \over 2} e^{-2\sqrt 2 r}|.
\end{equation}
Since both the forms of $V(r)$ and $r^*$ are very complicated, we are far
from obtaining
the exact form of $V(r^*)$. Instead we can find an approximate form.
{}From (19), in the asymptotically flat region one finds that $r^* \simeq r$.
(18) takes
the asymptotic form
\begin{equation}
V_{r*\to \infty} \simeq 2 \exp( -2 \sqrt 2 r^*).
\end{equation}
On the other hand, near the horizon ($r=r_o$) one has
\begin{equation}
r^* \simeq - { 1 \over 2 \sqrt 2( 1 - e^{ - 2 \sqrt 2 r})}.
\end{equation}
Approaching the horizon $(r\to r_o, r^* \to -\infty)$, the potential takes
the form
\begin{equation}
V_{r*\to -\infty}= - {1 \over 4 r^{*2}}.
\end{equation}
Using (20) and (22) one can construct the approximate form $V_{app}(r^*)$
(Fig. 2).
This is also a double-humped barrier well which is  localized at the origin
of $r^*$.
Our stability analysis is  based on the equation
\begin{equation}
\{ {d^2 \over dr^{*2}} + \omega^2 - V_{app}(r^*)\} \tilde t = 0.
\end{equation}

As we have probably guessed, two kinds of solutions to the Schr\"odinger
equation correspond to
 the bound and scattering states. In our case $V_{app}(r^*)$ admits  two
solutions  depending on
the signs of the energy.

(i) For $E>0(\omega=$ real), the asymptotic solution for $\tilde t$ is given by

\begin{eqnarray}
\tilde t_\infty & = & \exp(i \omega r^*)  + R \exp(- i \omega r^*)~~~~~~~~
( r^*  \to \infty ),  \\
\tilde t_{EH} & = & T\exp( i\omega r^*)~~~~~~~~~~~~~~~~~~~~~~~~~~~
( r^*  \to - \infty ),
\end{eqnarray}
where $R$ and $T$ are the scattering amplitudes of two waves which are
reflected and transmitted by the potential $V_{app}(r^*)$, when a wave of unit
amplitude with the frequency $\omega$ is incident on the black hole from
infinity.

\noindent $(ii)$ For $E<0(\omega =-i \alpha$, $\alpha$ is positive and real),
 we have the bound state.
Eq. (23) and possible asymptotic solutions are given by

\begin{eqnarray}
{d^2 \over d r^{*2}}\tilde t & = & (\alpha^2 + V_{app}(r^*)) \tilde t,     \\
\tilde t_\infty & \sim & \exp(\pm \alpha r^*),~~~~~~~~ ( r^*  \to \infty )  \\
\tilde t_{EH}   & \sim & \exp(\pm \alpha r^*) ~~~~~~~ ( r^*  \to - \infty ).
\end{eqnarray}
To ensure that the perturbation falls off to zero for large $r^*$, we choose
$\tilde t_\infty \sim \exp (-\alpha r^*)$.  In the case of $\tilde t_{EH}$,
the solution
$\exp (\alpha r^*)$ goes to zero as $r^* \to - \infty$.
Now let us observe whether or not $\tilde t_{EH} \sim \exp (\alpha r^*)$
can be matched
to $\tilde t_\infty \sim \exp (-\alpha r^*)$.
Assuming $\tilde t$ to be positive,
the sign of $d^2 \tilde t / dr^{*2}$
can be changed from $+$ to $-$ as  $r^*$
goes from $\infty$ to $-\infty$.
If we are to connect $\tilde t_{EH}$ at one end to a decreasing solution
$\tilde t_\infty$
at the other, there must be a point ($d^2\tilde t/ dr^{*2}<0$,
$d \tilde t/dr^*=0$) at which the signs of $\tilde t$ and
$d^2\tilde t/dr^{*2}$ are opposite : this is  compatible with  the shape of
$V_{app}(r^*)$ in Fig.2. It thus is possible for
$\tilde t_{EH}$ to be connected to $\tilde t_\infty$ smoothly.  Therefore a
bound state solution
is given by

\begin{eqnarray}
\tilde t_\infty & \sim & \exp(- \alpha r^*),~~~~~~~~ ( r^*  \to \infty )     \\
\tilde t_{EH}   & \sim & \exp( \alpha r^*) ~~~~~~~~ ( r^*  \to - \infty ).
\end{eqnarray}
This is a regular solution everywhere in space at the initial time $t=0$.
However, $\omega=-i\alpha$ implies
$t_\infty (r^*, t)=\tilde t_\infty(r^*)\exp(-i\omega t) \sim \exp (-\alpha
r^*) \exp (\alpha t)$ and
$t_{EH} (r^*,t)=\tilde t_{EH}(r^*) \exp(-i\omega t) \sim \exp (\alpha r^*)
\exp (\alpha t)$.
This means
that there exists an exponentially growing mode with time.
Therefore, the 2D extremal ground state  is
classically unstable. The origin of this instability comes from  a
double-humped barrier well.
This potential appears when the nonextremal black hole approaches the
extremal limit.
As is discussed in Ref. [8], the quantum stress tensor of a scalar field
(instead of the tachyon)
in the extremal black hole
diverges at the horizon. This means that the extremal black hole is
quantum-mechanically unstable.
This divergence can be better understood by the regarding an extremal black
hole as the limit of a
nonextremal one. A nonextremal black hole has an outer (event) and an inner
(Cauchy) horizon,
 and these come together
in the extremal limit.  In this case, we find that if we adjust  the
quantum state of the scalar
 field so that  the stress tensor is finite at the outer horizon, it always
diverges at
the inner horizon.
Thus it is not so surprising that in the extremal limit ( when the two
horizons come together)
the divergence persists, although it has a softened form.
By the similar way, it conjectures that the classical instability
originates from the instability
(blueshift) of the inner horizon [7].

In conclusion, the 2D extremal electrically charged black hole cannot be a
candidate for the stable endpoint
of the Hawking evaporation.

\acknowledgments

This work was supported in part by the Basic Science Research Institute
Program,
Ministry of Education, Project No. BSRI-95-2413
and by Nondirected Research Fund, Korea Research Foundation, 1994.

\figure{ Fig.1 : The $M=Q= \sqrt 2$ graph of the effective potential of
tachyon  ($ V(r)$).
     The event horizon is at $r_o= -0.245$. This takes a
   double-humped barrier well outside the black hole.

\figure{ Fig.2 :The approximate  form of potential($V_{app}(r^*)$)  outside
black hole.
This also takes a double-humped barrier well. This is localized at $r^*=0$,
falls to zero
exponentially as $r^* \to \infty$ and inverse-squarely as $r^* \to -\infty$
(solid lines).
 The dotted line is used to connect two boundaries.}
\end{document}